# Quantum Computing using Linear Optics

Todd B. Pittman, Bryan C. Jacobs, and James D. Franson

Quantum computers are expected to be able to solve mathematical problems that cannot be solved using conventional computers. Many of these problems are of practical importance, especially in the areas of cryptography and secure communications. APL is developing an optical approach to quantum computing in which the bits, or "qubits", are represented by single photons. Our approach allows the use of ordinary (linear) optical elements that are available for the most part as off-the-shelf components. Recent experimental demonstrations of a variety of logic gates for single photons, a prototype memory device, and other devices will be described.

## 1. Introduction

In the early 1980's, Richard Feynman showed that there are fundamental limitations in trying to perform simulations of complex quantum systems on conventional computers, regardless of their size or speed [1]. He noted, however, that these problems could be overcome, at least in principle, by building computers based on quantum mechanics instead of classical physics. One naturally wondered if these "quantum computers" would be useful for other applications, assuming that they could eventually be built.

The answer was shown to be "yes" when, nearly a decade later, Peter Shor discovered a quantum computing algorithm [2] for efficiently factoring large integers – a problem that has no efficient solution on conventional computers and forms the basis of many secure communications protocols. Soon thereafter, it was shown that a quantum computer could also be used to search an unstructured database much faster than any conventional computer [3]. Due to these critical theoretical developments, there has been a recent explosion of experimental work aimed towards building a quantum computer. Researchers in many different areas of physics are actively pursuing a variety of methods to accomplish this challenging goal.

APL is currently developing an optical approach to quantum computing. In our approach the quantum bits, or "qubits", of information are represented by the quantum state of single photons. For example, the logical value 0 can be represented by a horizontally polarized photon, while the logical value 1 can be represented by a vertically polarized photon. Alternatively, 0 and 1 could be represented by the presence of a single photon in one of two optical fibers (see Fig.1). As discussed in an earlier *Technical Digest* article [4], the power of quantum computing is related to the fact that quantum mechanics allows the qubits to be in states that do not correspond to specific values of 0 or 1. For example, a so-called superposition state of the form

$$|\psi\rangle = \alpha|0\rangle + \beta|1\rangle \tag{1}$$

is characterized by two complex numbers, α and β, whose squares correspond to the probability of measuring a value of 0 or 1. In stark contrast to classical bits (which always have a definite value of either 0 or 1) the qubits can, in some sense, behave as if they had the values of 0 and 1 at the same time. In addition to photons, there are many other physical quantum systems being considered for use as qubits. For example, a single two-level atom in its ground state could correspond to a 0, while the same atom in its excited state would correspond to a 1. Research along these lines is actively being pursued within the context of ion-trap [5] and nuclear magnetic resonance (NMR) [6] approaches.

## 2. Linear optics quantum computing

The primary advantage of an optical approach to quantum computing is that it would allow quantum logic gates and quantum memory devices to be easily connected together using optical fibers or wave-guides in analogy with the wires of a conventional computer. This affords a type of modularity that is not readily available in other approaches. For example, the transfer of qubits from one location to another in ion-trap or NMR systems is a very complex process.

The main drawback to an optical approach has been the implementation of the quantum logic gates needed to perform calculations. An important example of a quantum logic gate is the so-called controlled-NOT (CNOT) gate, which has been shown to be a universal gate for quantum computers in the same way that the classical NAND gate is a universal gate for conventional computers [7]. As described in reference [4], a CNOT gate has two inputs (a control qubit and a target qubit) and operates in such a way that the NOT operation (bit flip) is applied to the target qubit, provided the control qubit has a logical value of 1.

Such a logic operation is inherently nonlinear because the state of one quantum particle must be able to control the state of the other. In an optical approach, this is equivalent to requiring a nonlinear interaction between two single photons, which is typically an extremely weak effect. Conventional nonlinear optical effects, such as frequency doubling of a light beam, are usually only observed in experiments involving intense laser pulses containing billions of photons [8]. Although several ingenious methods for producing nonlinear interactions at single-photon intensity levels have been considered, they are thought to be either too weak [9] or accompanied by too much loss [10] to be of use for practical quantum CNOT gates.

It has recently been shown, however, that near-perfect optical quantum logic gates, such as a CNOT gate, can be implemented without the need for a nonlinear interaction between two single photons [11]. Logic gates of this kind can be constructed using only linear optical elements, such as mirrors and beam splitters, additional resource photons, and triggering signals from single-photon detectors. In this "linear optics quantum computing" (LOQC) approach, the required nonlinearity arises from the quantum measurement process associated with the detection of the additional resource photons [11]. Roughly speaking, a single-photon detector either goes off or not, which is a very nonlinear response.

The basic idea of a LOQC-type CNOT gate is illustrated in Figure 2. In addition to the control and target photons, the additional photons (called ancilla) are injected into a "black-box" containing only linear optical elements. The optics are designed in such a way that there are three types of outcomes from the device, each of which is signaled by a unique combination of triggering events at a series of single-photon detectors. In one set of outcomes, we know that the control and target photons are in the desired logical output state. In the second type of outcome, the control and target are known to be in the wrong output state, but they can be corrected in a known way using real-time corrections known as feed-forward control [12]. The third type of outcome indicates that the control and target photons have been lost, or are in a logical state that cannot be corrected.

These LOQC logic gates are referred to as probabilistic devices because they occasionally fail, but it is known when a failure has occurred. In addition, the gates can be designed in such a way that the probability $P_f$ of a failure event can be made arbitrarily small. In the original LOQC proposal [11], it was shown that $P_f$ can be proportional to 1/N, where N is the number of ancilla photons consumed by the gate. In a subsequent paper [13], we described an alternative approach in which $P_f$ scales as $1/N^2$, which greatly reduces the resources required for a given gate fidelity.

### 3. Experimental quantum logic gates

Our goal was to design LOQC logic devices that are as simple, stable, and robust as possible. We therefore use qubits represented by the polarization states of single photons, as illustrated in Fig. 1a. Polarization encoded qubits are more resistant to certain kinds of experimental errors, and they are easier to manipulate than the "path encoded" qubits of Fig. 1b.

The use of polarization based qubits allowed us to design a CNOT gate using only two polarizing beam splitters, two polarization sensitive detectors, and two ancilla photons, as shown in Fig. 3 [14]. In this device, the two ancilla photons are in a quantum-mechanically correlated or "entangled" state described by:

$$|\psi\rangle = \frac{1}{\sqrt{2}}(|0,0\rangle + |1,1\rangle) \qquad (2)$$

In such a quantum state, the logical value (i.e., polarization) of each of the ancilla photons is totally undefined, but measurements will always find the logical values of the two photons to be the same. Quantum-mechanical correlations of this kind are stronger than those allowed by classical physics, and the possibility of their existence prompted

Einstein to question the completeness of quantum mechanics in the early days of the theory [15]. Nonetheless, advances in modern technology have allowed entangled states of this kind to be produced and measured in the laboratory [16]. The device shown in Figure 3 exploits the entanglement of the ancilla pair to implement the desired CNOT logic operation on the input control and target qubits. The correct logical output is known to have been produced whenever each of the detectors registers one and only one photon, which occurs with a probability of ¼ [14].

We recently performed an experimental demonstration of a CNOT gate of the kind shown in Fig. 3 [17]. In our experiment, the arrangement of the polarizing beam splitters was altered so that the role of the entangled photon pair could be replaced by a single ancilla photon propagating through the entire device. This simplified the technical requirements of the experiment by reducing the total number of photons involved from 4 to 3. However, in this simplified configuration the operation of the logic gate could only be verified by measuring (and thus destroying) the control and target qubits after they exited the device. Nonetheless, the experimental results represented the first demonstration of a CNOT gate for single photons, and a tangible step towards full-scale quantum computing using linear optics.

A photograph of the experimental apparatus is shown in Fig. 4. In this proof-of-principle experiment, one of the three required photons was obtained from an attenuated laser pulse, while the other two were produced through the process of spontaneous parametric down-conversion (SPDC) [18]. The SPDC process consists of passing an intense laser pumping pulse through a nonlinear medium in such a way that correlated pairs of photons occasionally emerge. SPDC is a purely quantum-mechanical phenomenon that can be viewed as the annihilation of a single photon in the laser pulse followed by the creation of two lower energy down-conversion photons, under the conditions that energy and momentum are conserved.

The operation of our CNOT logic gate relied on multi-photon quantum interference effects that required the three photons to be completely indistinguishable, aside from their polarizations (i.e., logical values).  This required a combination of precise spectral filtering, the use of single-mode optical fibers for spatial mode-matching, and timing precision on the order $10^{-13}$ seconds.  Once these parameters were optimized, the CNOT gate could be tested using polarizing optics to control the values of the control and target input qubits, and by using polarization analyzers followed by single-photon detectors to test and measure the output of the device.

An example of these types of measurements is shown in Fig. 5.  In comparison with the theory, the experimental results clearly demonstrate the desired logical truth table of a CNOT gate aside from technical errors on the order of 10%.  Experiments aimed at reducing these technical errors, as well as utilizing an entangled ancilla pair of resource photons as shown in Fig. 3, are currently underway in the laboratory.

In addition to a CNOT gate, we have also demonstrated a number of other important LOQC gates, including a quantum parity check, an exclusive-OR gate [19], and a quantum encoding device [20] capable of encoding (copying) the value of a single qubit into a logical state represented by two photons.  This encoding operation can be used to provide redundancy that can protect against photon loss or other errors that may occur in a realistic environment.  In addition to its use in quantum computing applications, an encoder of this kind can be used to implement a "quantum relay" that would help extend the range of quantum cryptography systems [21].

**4. Single-photon sources**

All of the logic gates describe above rely on small numbers of ancilla photons, and typically operate with probabilities of success in the range of ¼ to ½. Although these gates have a number of important applications, full-scale quantum computers will require highly-efficient gates involving larger numbers of ancilla photons [22]. As a result, the development of reliable sources capable of emitting a single photon at well-defined time intervals is a key ingredient to the realization of LOQC.

The development of a source of single photons is a critical requirement and it cannot be accomplished by re-engineering a conventional light source. For example, the number of photons in a laser pulse follows a Poisson distribution. Attenuating a train of laser pulses until each pulse contains an average of one photon will result in a few pulses that actually contain zero photons or more than one. Injecting those pulses into an LOQC logic gate would produce undesirable errors. Although quantum error correction techniques that are analogous to classical parity checks do exist, it is necessary to keep the intrinsic error rate below a certain threshold.

One method for realizing a true single-photon source is through spontaneous emission from an isolated two-level quantum system. For example, a single two-level atom in its excited state can emit only one photon, as was experimentally verified nearly 30 years ago [23]. In recent years, the suitability of a variety of other single-photon emitters have been investigated, including single molecules, quantum dots, and solid-state defects such as color centers [24].

In contrast to these approaches, APL's contribution has been the development of a single-photon source based on the photon pairs produced in SPDC [25]. The basic idea of this source is illustrated in Fig. 6. Because SPDC is known to produce pairs of photons, the detection of one photon of a pair can be used to signal the presence of the twin photon. A high speed optical switch is then used to store the twin photon in a storage loop until it is needed, at which time is can be switched back out of the storage loop. We have experimentally demonstrated a single-photon source of this kind. Its performance is currently limited by losses in the optical switch, but we are now developing a low-loss switch for single photons. Storage loops and low-loss switches can also form the basis of a quantum memory device for single-photon qubits [26].

## 6. Quantum Circuits

As we mentioned earlier, one of the main advantages of an optical approach to quantum computing is the ability to connect logic and memory devices using optical fibers, in analogy with the use of wires in conventional electronic circuits. In order to demonstrate this capability, we recently constructed and tested a relatively simple quantum circuit that combines two linear optics quantum logic gates [27] to perform a useful function. The circuit consisted of two probabilistic exclusive-OR (XOR) gates in series, as shown in Figure 7. Two single-photon qubits formed the input to the first XOR gate. The output of that gate then served as the input to second XOR gate, along with a third single-photon qubit.

Because an XOR gate can be used to measure the parity of its inputs, it can be shown that the circuit of Fig. 7 calculates the parity of the three input qubits when they all have specific logical values of 0 or 1. When one or more of the input qubits is in the more general superposition state of equation (1), the circuit of Fig. 7 produces a quantum mechanical output state that can not be reproduced by any classical device. We have demonstrated both the classical truth table corresponding to well-defined input values, as well as a variety of quantum interference effects associated with superposition states. Although XOR gates are not reversible, this simple circuit is still useful in a number of important applications and it demonstrates our ability to connect independent devices using optical fibers to form more complex circuits.

## 7. Summary

In the past few years, the prospects for quantum computing have progressed from a fascinating academic exercise to one with important applications. A linear optics approach appears to be a promising method for eventually building a full-scale quantum computer. APL has been actively developing and demonstrating many of the basic building blocks that are required, including the first CNOT gate for single-photon qubits [17], as well as a prototype single-photon source [25], a quantum memory device [26], and a photon-number resolving detector [28]. In addition, we recently demonstrated the first quantum circuit for photonic qubits [27].

Although these proof-of-principle experiments are encouraging, many significant technical challenges remain. In particular, the need for large numbers of ancilla photons to achieve low error rates may increase the complexity of such a quantum computer and increase the amount of resources required. With that in mind, we are investigating a variety of ways to reduce the dependence on large numbers of ancilla photons. For example, we have recently shown that the quantum Zeno effect can be used to completely suppress the failure events that would otherwise occur in these probabilistic quantum logic gates [29]. This new approach would eliminate the need for ancilla photons and make LOQC more practical for large-scale applications. We are planning to perform experiments of that kind in the near future.

In summary, the development of quantum computers would have a major impact in a number of areas of practical importance, including cryptography, secure communications, and optimization problems. APL has demonstrated the basic building blocks of an optical approach to quantum computing. Although a number of challenges remain, our optical approach appears promising as a method for eventually building a full-scale quantum computer.

**Acknowledgements**

This work was supported by ARO, ARDA, ONR, and IR&D funding.

**Biographies**

Todd B. Pittman received a B.S. in physics from Bucknell University in 1990, and a Ph.D. in physics from the University of Maryland Baltimore County in 1996. He joined the Milton S. Eisenhower Research and Technology Development Center at APL in 1996, where he is now a member of the Senior Professional Staff. His current research activities include quantum and nonlinear optics, and quantum information processing. His e-mail address is todd.pittman@jhuapl.edu

Bryan C. Jacobs received a B.S. in electrical engineering from Drexel University in 1989, an M.S. in applied physics from the Johns Hopkins University in 1994, and a Ph.D. in physics from the University of Maryland Baltimore County in 2002. He joined the Strategic Systems Department at APL in 1989 and transferred to the Milton S. Eisenhower Research and Technology Development Center in 1996, where he is now a member of the Senior Professional Staff. He has collaborated on several basic research and development projects, including the Advanced Inertial Sensors Project, numerous quantum cryptography projects, and quantum computing research. His e-mail address is bryan.jacobs@jhuapl.edu

James D. Franson received a B.S. in physics from Purdue University in 1970 and a Ph.D. in physics from the California Institute of Technology in 1977. He joined the Strategic Systems Department at APL in 1978 and transferred to the Milton S. Eisenhower Research and Technology Development Center in 1996. He is a member of the Principal Professional Staff and a Fellow of the Optical Society of America. His current research activities include quantum information processing and the foundations of quantum mechanics. His e-mail address is james.franson@jhuapl.edu

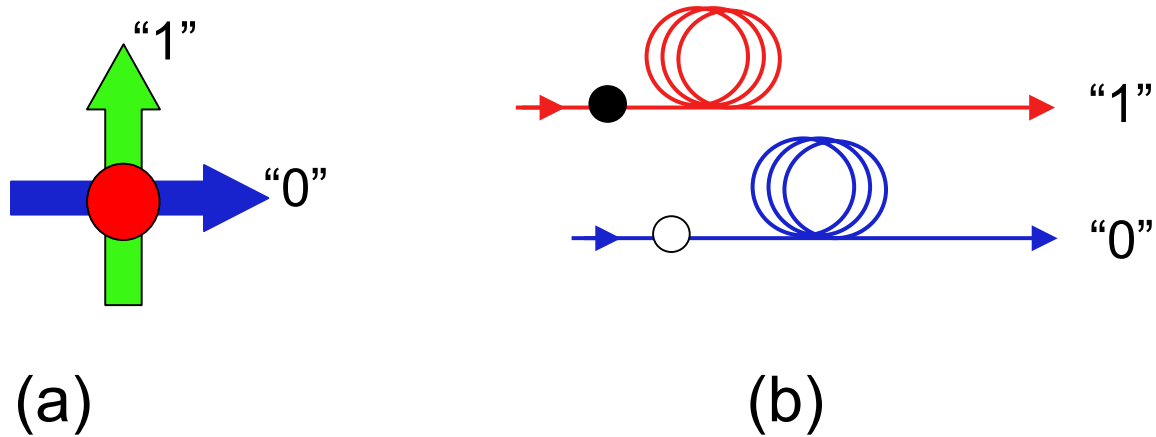

Figure 1: Two methods for implementing quantum bits, or "qubits", using the quantum states of single photons [4]. (a) Polarization encoding in which a horizontally polarized single photon represents a logical value 0 and a vertically polarized single photon represents a logical value 1. (b) Path encoding, where the presence of a single photon in one of two optical fibers represents a logical value of 0 or 1.

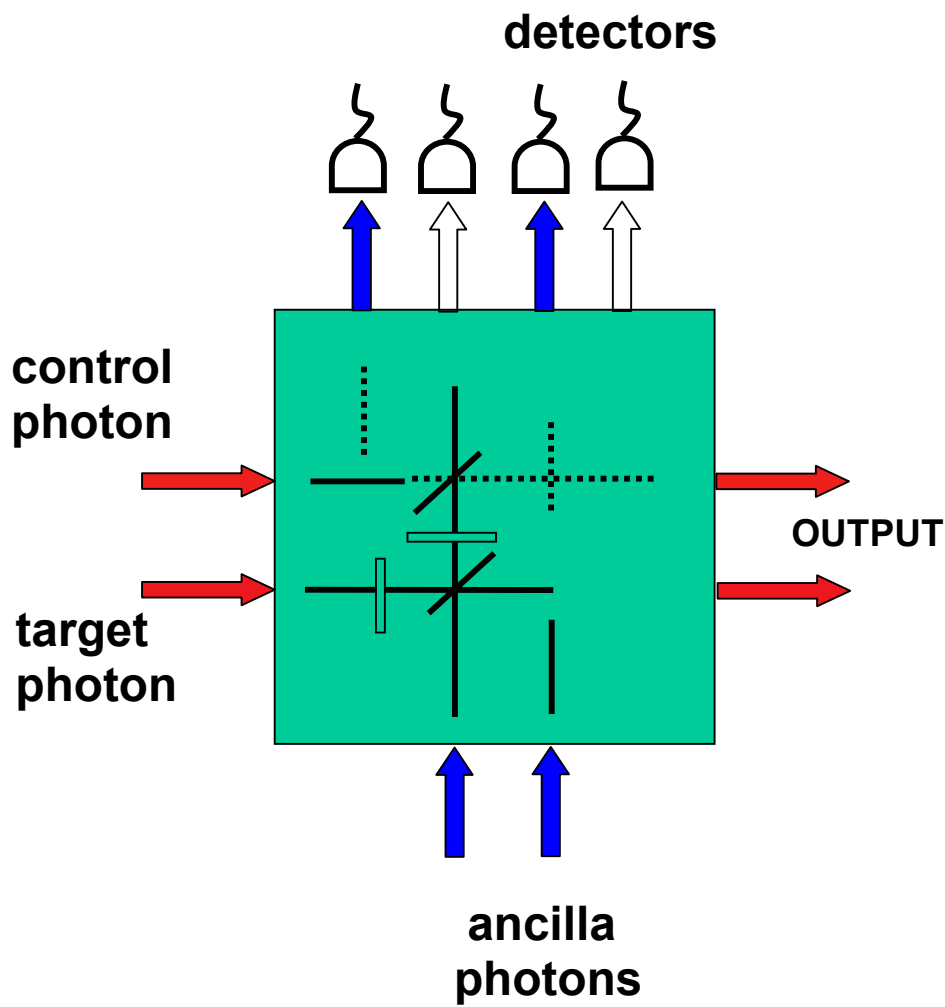

Figure 2. Basic idea of a two-input quantum logic gate constructed using linear optical elements, additional resource photons (ancilla), and single-photon detectors. The ancilla photons are combined with the logical qubits using linear elements, such as beam splitters and phase shifters. The quantum state of the ancilla photons is measured after they leave the device. The correct logical output is known to have been produced when measurements on the ancilla photons produce certain results. The output can be corrected for other measurement results.

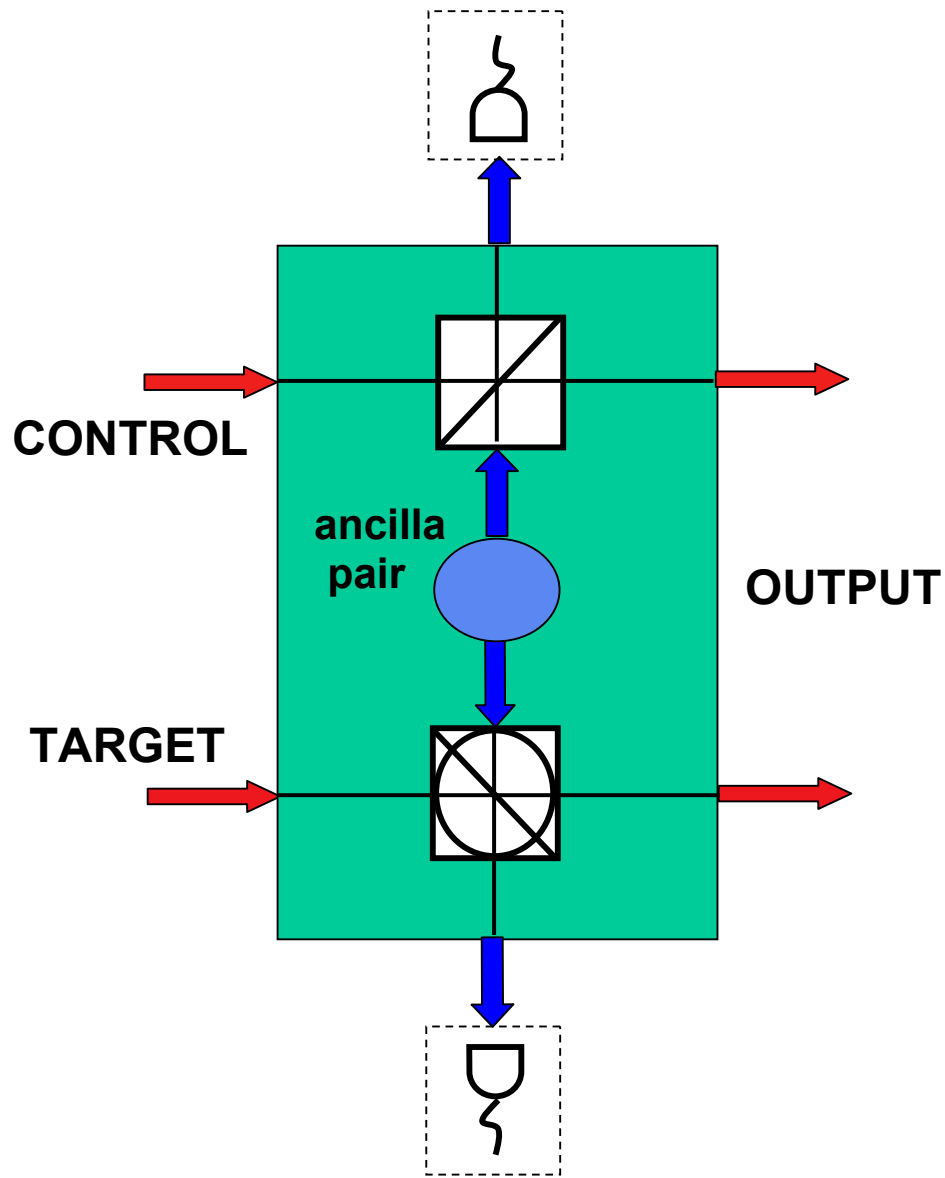

Figure 3. Overview of the APL linear optics quantum controlled-NOT gate [14]. In addition to the input control and target qubits, the device consists of two polarizing beam splitters, an entangled pair of resource ancilla photons, and two single-photon detectors. The output is known to be correct when each of these detectors registers one photon, which occurs with a probability of ¼.

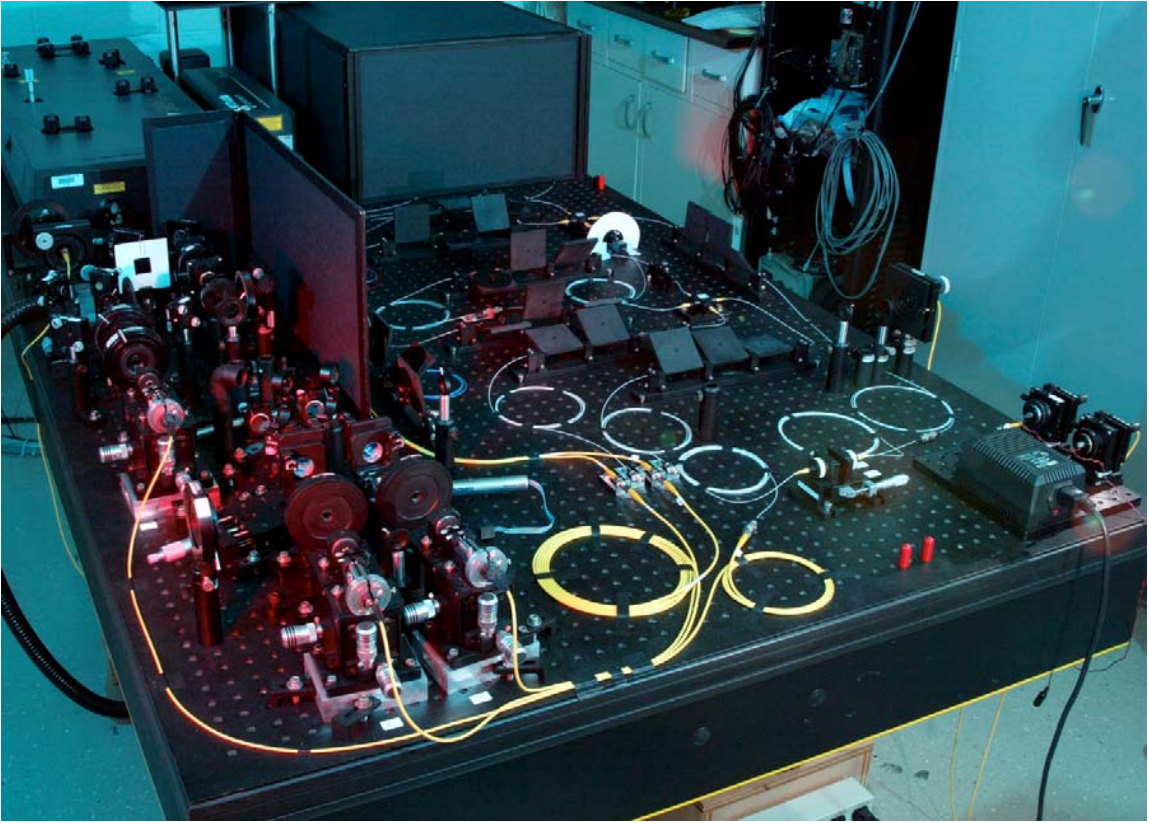

Figure 4. Photograph of the experimental apparatus used to demonstrate the first quantum controlled-NOT logic gate for single-photon qubits. The experimental components included a mode-locked Ti-Sa. laser, single-mode fiber components, parametric down-conversion photon sources, and low-noise single-photon detectors.

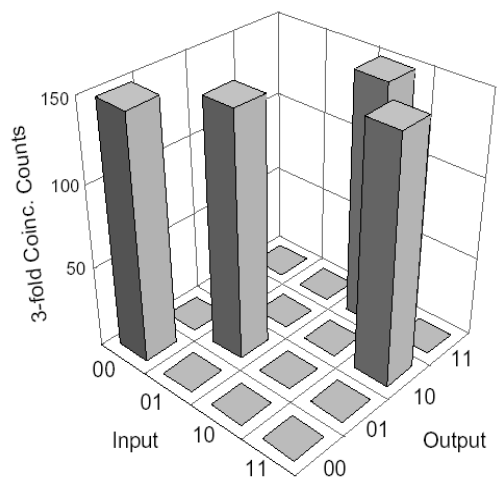 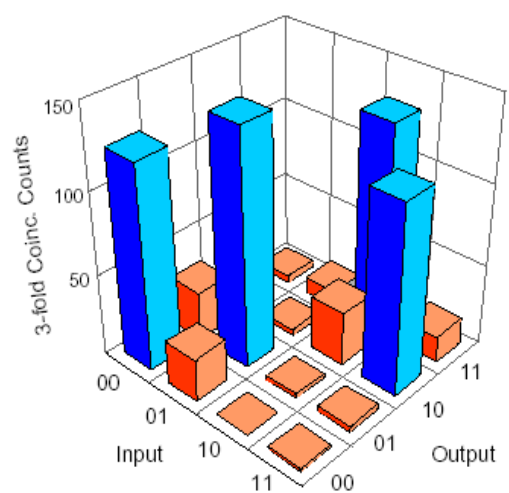

Figure 5. Experimental results demonstrating the logical truth table for the first quantum controlled-NOT gate [17] for single photons. The NOT operation (bit flip) is applied to the target qubit if and only if the control qubit has the logical value 1. The experimental data is seen to be in agreement with the theory, aside from average technical errors on the order of 10%.

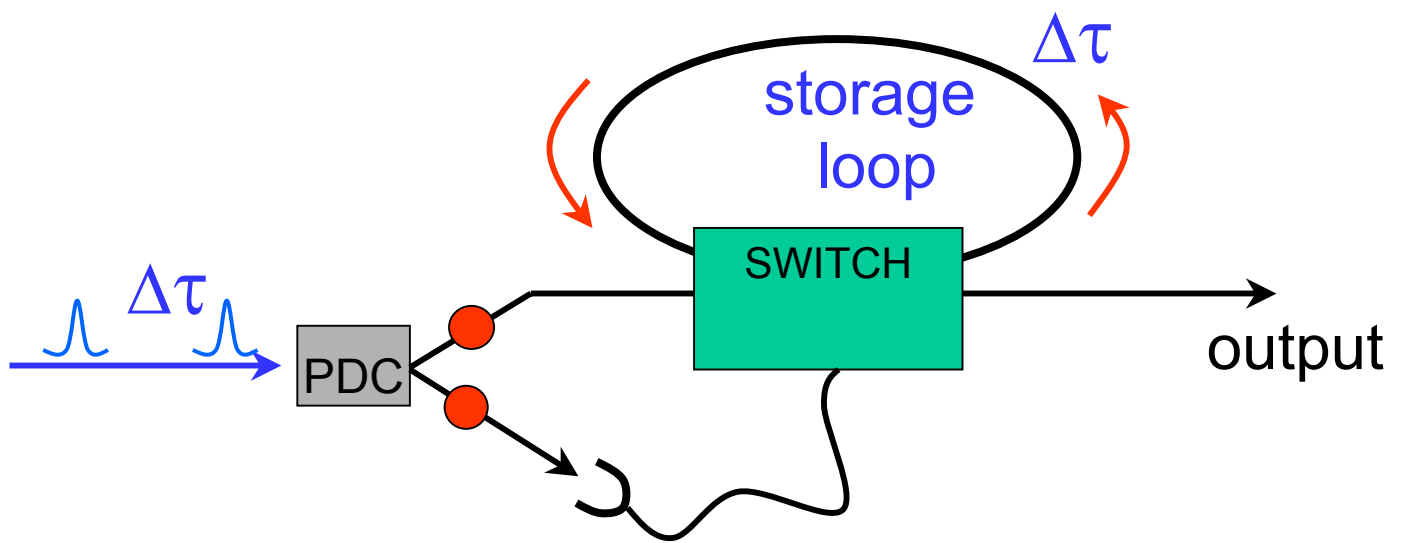

Figure 6. Overview of APL's single-photon source [25]. A parametric down-conversion crystal (PDC) is pumped by a train of laser pulses separated in time by $\Delta\tau$, which causes it to randomly emit pairs of correlated photons [18]. Once a pair is emitted, the detection of one of the photons activates an electro-optic switch that is used to route the other photon into a storage loop. The stored photon can then be switched back out of the loop when it is needed.

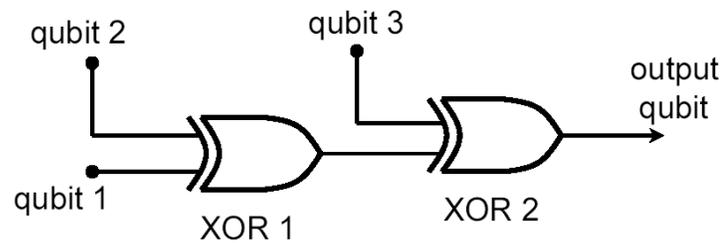

Figure 7: A simple logic circuit for photonic qubits.  This circuit calculates the parity of three input qubits using two exclusive-OR (XOR) logic gates.  We have demonstrated the classical truth table for this circuit along with a variety of quantum interference effects when the input qubits are in quantum superposition states.